\documentclass[a4paper,fleqn,usenatbib]{mnras} 

\usepackage{mathptmx}
\usepackage{xspace}
\usepackage[T1]{fontenc}
\usepackage{ae,aecompl}

\usepackage{graphicx}	% Including figure files
\usepackage{amsmath}	% Advanced maths commands
\usepackage{amssymb}	% Extra maths symbols
\usepackage{subfigure}

\usepackage{aasmacros}
\usepackage{units}% To enter units like \unit{MeV} or \unit[1]{km/sec}

\newcommand{\Msun}{\mathrm{M_{\sun}}}

\newcommand{\kms}{$\rmn{km\,s^{-1}}$} 
\newcommand{\vmax}{$V_{\rmn{max}}$\xspace}

\def\gsim{ \lower .75ex \hbox{$\sim$} \llap{\raise .27ex \hbox{$>$}} }
\def\lsim{ \lower .75ex \hbox{$\sim$} \llap{\raise .27ex \hbox{$<$}} }

\title[TBTF with baryons and sterile neutrinos]{Addressing the too big to fail problem with baryon physics and sterile neutrino dark matter}
\author[M. R. Lovell et al.]{Mark R. Lovell$^{1,2,3}$ \thanks{E-mail:
    lovell@mpia-hd.mpg.de}, Violeta Gonzalez-Perez$^{4}$, Sownak Bose$^{5}$, Alexey Boyarsky$^{2}$,\newauthor Shaun Cole$^{5}$, Carlos S. Frenk$^{5}$, and Oleg Ruchayskiy$^{6}$\\
    $^{1}$GRAPPA, Universiteit van Amsterdam, Science Park 904, 1098 XH
Amsterdam, The Netherlands\\
$^{2}$Instituut-Lorentz for Theoretical Physics, Niels Bohrweg 2, 2333 CA Leiden, The Netherlands\\
$^{3}$Max-Planck-Institut f\"ur Astronomie, K\"onigstuhl 17, D-69117 Heidelberg, Germany\\
$^{4}$Institute of Cosmology and Gravitation, University of Portsmouth, Dennis Sciama Building, Portsmouth PO1 3FX, UK\\
$^{5}$Institute for Computational Cosmology, Durham University, South Road, Durham, UK, DH1 3LE\\
$^{6}$Discovery Center, Niels Bohr Institute, Blegdamsvej 17, 2100, Copenhagen, Denmark}

\date{Accepted *** Received ***; in original
  form ***} 
\pubyear{2015}

\begin{document}

  \label{first page}
  \pagerange{\pageref{firstpage}--\pageref{lastpage}} 
  
  \maketitle

  \begin{abstract}
    N-body dark matter simulations of structure formation in the $\Lambda$CDM model predict a population of subhalos within
    Galactic halos that have higher central densities than inferred
    for the Milky Way satellites, a tension known as the `too big
    to fail' problem. Proposed solutions include baryonic effects, a
    smaller mass for the Milky Way halo, and warm dark matter. We test
    these possibilities using a semi-analytic model of galaxy
    formation to generate luminosity functions for Milky Way
    halo-analogue satellite populations, the results of which are then
    coupled to the Jiang \& van den Bosch model of subhalo stripping
    to predict the subhalo \vmax functions for the 10 brightest
    satellites. We find that selecting the brightest satellites (as
    opposed to the most massive) and modelling the expulsion of gas by
    supernovae at early times increases the likelihood of generating
    the observed Milky Way satellite \vmax function. The preferred halo mass is
    $6\times10^{11}\Msun$, which has a 14~per~cent probability to host
    a \vmax function like that of the Milky Way satellites. We
    conclude that the Milky Way satellite \vmax function is
    compatible with a CDM cosmology, as previously found by Sawala et
    al. using hydrodynamic simulations. Sterile neutrino-warm dark matter
    models achieve a higher degree of agreement with the observations,
    with a maximum 50~per~cent chance of generating the observed Milky
    Way satellite \vmax function. However, more work is required to check
    that the semi-analytic stripping model is calibrated correctly for each sterile neutrino cosmology.

  \end{abstract}


  \section{Introduction}
  \label{intro}

The properties of the satellite galaxies of the Milky Way offer an
  opportunity to study the process of galaxy formation and the nature
  of dark matter. They are among the intrinsically faintest galaxies
  that have been observed, and thus constitute an `extreme laboratory'
  in which to examine the interplay between the underlying
  cosmological model and astrophysical processes. One property that
  has been of particular interest is the central density of these
  objects. The likely distribution of densities -- or the more
  observationally accessible central velocity dispersions -- can be
  predicted from simulations of Milky Way-analogue systems using a
  combination of the satellites' density profiles and mass functions.
% A successful model of galaxy formation will need to produce a
%distribution of satellite central densities that match those
 % inferred from observations of the Milky Way satellites.

  The ability to compare theoretical predictions with observational
  measurements was made possible by two, almost simultaneous
  developments. First, simulations of Milky Way-analogue haloes
  achieved sufficient spatial resolution to resolve the properties of
  cold dark matter (CDM) subhaloes on scales of $\sim$100~pc
  \citep{Diemand07,Springel08b}, which is smaller than the size of the
  brightest satellite galaxies. These simulations predicted that the
  satellites had cuspy density profiles, and that these profiles
  were better described by the Einasto profile \citep{Navarro10} than
  the $\sim r^{-1}$ profile predicted for isolated
  haloes \citep{NFW_96,NFW_97}. Second, masses within the half-light
  radii of the Milky Way satellites were estimated using the methods
  developed by \citet{Walker09,Walker10} and \citet{Wolf10} (but see
  \citealp{Campbell16} for a realistic estimate of the errors). The
  results of these observations were interpreted by \citet{Walker11}
  and \citet{Gilmore2007} as evidence that the satellites had cored,
  rather than cuspy, profiles and were thus in tension with the CDM
  simulation results. However, this interpretation remains contentious
  \citep{Strigari10,Strigari14}.

  A second tension between the observations and the theoretical
  predictions concerns the expected abundances of dense, massive
  ($>10^{10}\Msun$) dark matter subhaloes around Milky Way hosts.
  \citet{BoylanKolchin11,BoylanKolchin12} found that the six Milky
  Way-analogue dark matter simulations of the Aquarius project
  predicted a population of subhaloes that were too dense and massive
  to host the brightest observed satellites. This problem was first identified
  by inferring the central densities of simulated subhaloes from the
  peak of their circular velocity curves, denoted \vmax
  \citep{BoylanKolchin11}, and persisted when the highest resolution
  simulations were used to measure the central densities directly
  \citep{BoylanKolchin12}. The masses of these simulated subhaloes
  were large enough to guarantee that gas should collapse within them
  and form a comparatively bright satellite galaxy that should have
  been detected by satellite galaxy searches \citep{Parry2012}. This
  issue became known as the `too big to fail' problem.

  Proposed solutions to this problem have adopted at least one out of
  two approaches. The first is to decrease the number of massive
  satellites around the Milky Way. This has been achieved for
  satellites with \vmax$>30$~\kms by invoking a relatively low mass
  for the
  Milky Way halo \citep{Wang12,Cautun14b}; less massive
  (\vmax$<20$~\kms) subhaloes are prevented from forming galaxies
  by reionisation and supernova feedback
  \citep[e.g.][]{Bullock_00,Benson_02, Sawala16b}. These models also
  predict scatter in the luminosity-mass relation of galaxies; thus,
  the brightest galaxies need not reside in the most massive haloes,
  as seen in observations \citep{GuoQi15}.

  The second approach is to appeal to baryon effects. One possibility
  is that adiabatic contraction of the gas initially draws dark matter
  to the halo centre, only to be evacuated violently when supernova
  feedback occurs \citep{NEF96,Pontzen_Governato_11}, although if the
  feedback is too weak then adiabatic contraction of the gas can
  increase the density of simulated galaxies still further and thus
  make the discrepancy with observations even worse
  \citep{diCintio12}. Another possibility is that early feedback from
  reionisation and supernovae lowers the baryonic mass of the halo, so
  that less mass accretes onto the halo at later times and the
  redshift zero mass is smaller than in the pure N-body prediction
  \citep{Sawala_13,Sawala16a}. A third possibility is for tides to
  remove material from satellites \citep{Fattahi16}. In practice,
  these methods also reduce the total mass of the satellites and can
  bias galaxy formation efficiency such that some late-forming massive
  subhaloes host relatively faint galaxies \citep{Sawala16a}. A fourth
  solution, which affects both the abundance and density of the most
  massive haloes, is a revision of the cosmological parameters.
  \citet{Polisensky14} argued that better agreement with observations
  was achieved with the
 cosmological parameters from the {\it Wilkinson Microwave Anisotropy
   Probe} (WMAP) 3-year results than with the 1-year values used in
 the original Aquarius simulations, due to the lower value of the
 power spectrum normalisation, $\sigma_8$. The result is that
 gravitational collapse begins at an epoch when the mean density of
 the Universe is lower.
 
An alternative set of solutions invokes alterations to the dark matter model. It has been shown that a velocity dependent self-interacting dark matter model (vdSIDM) successfully evacuates the right amount of dark matter from the subhalo centre, even creating a core as suggested by \citet{Walker11}, while remaining in agreement with bounds on dark matter self-interactions set by halo shapes \citep{Vogelsberger12,Zavala13,CyrRacine16,Vogelsberger16}. Another possibility is for the dark matter to interact with radiation, which also dilutes the central dark matter density \citep{Schewtschenko15}. A third possibility is that the dark matter is a warm dark matter (WDM) particle, such as the resonantly-produced sterile neutrino \citep{Dodelson94,Shi99,Dolgov:00,Asaka05} that may have already been detected in its X-ray decay channel \citep[e.g.][]{Bulbul14,Boyarsky14a}. WDM particles free stream out of small perturbations in the early Universe. This phenomenon reduces the abundance of $10^9-10^{10}\Msun$ haloes and delays the collapse of those that do form, to an epoch when the Universe is more diffuse and thus the haloes are less dense \citep{Lovell12}. The creation of a core due to primordial velocities does not help because  these are predicted to be smaller than $\sim1$~pc and therefore not relevant for the satellite internal kinematics \citep{Maccio12,Maccio13,Shao13}.        
 
The challenge of analysing all of these possibilities, some of which
are in competition and others complementary to one another, is
compounded by stochastic effects. Even within models restricted to CDM
which do not include baryonic processes large statistical
uncertainties are introduced by the stochastic formation of Milky
Way-like haloes and uncertainty in the Milky Way halo mass, which is
expected to be in the range $[0.5,2.0]\times10^{12}\Msun$
\citep{Kahn59, Sales07a,Sales07b,Li08,Busha11,Deason12,
  Wang12,Gonzalez13,BK13,Cautun14b,Penarrubia14,Piffl14,Wang15,Penarrubia16}.
In order to take account of these effects, \citet{JiangF15} computed
$\sim10000$ merger trees of Milky Way-analogue CDM haloes of a range
of masses using a Monte-Carlo (MC) method. They then used a semi-analytic
model of subhalo stripping \citep{JiangF16} to calculate the \vmax
functions of each halo realisation. They found the Milky Way system of
satellites, as defined by the inferred Milky Way satellite \vmax
function with \vmax$>15\rmn{kms}^{-1}$, to be a $\sim1$~per~cent
outlier of the MC-generated distributions.
  
  In this paper we also use a MC approach to investigate the \vmax function. Our method, however, differs from that of \citet{JiangF15} in many respects: 
  
  \begin{itemize}
  	\item We use the {\em ab initio} semi-analytic galaxy formation model,  {\sc galform } to populate haloes and subhaloes with galaxies. In this way, we can select satellites that are luminous, and in particular those with the highest luminosities.
	\item  We apply a correction for baryonic effects which changes the satellite \vmax values derived from hydrodynamical simulations. 
	\item We make use of new \vmax estimates for the Milky Way satellites based on the results of hydrodynamic numerical simulations \citep{Sawala16a}.
	\item We apply the method to a series of WDM models, specifically a range of sterile neutrino models whose decay is a plausible source of the recently discovered 3.5~keV line \citep[e.g.][]{Bulbul14, Boyarsky14a}.
   \end{itemize}
	
%	A thorough treatment of the sterile neutrino model would require calibration against N-body simulations, which is beyond the scope of this paper; our results for this model are therefore illustrative rather than precise predictions. 

   This paper is organised as follows. In Section~\ref{meth} we describe our methods. These include the generation of merger trees, the population of these merger trees with galaxies, the algorithm for comparing these galaxies with observations, and a discussion of the sterile neutrino models used. We present our results in Section~\ref{res}, and draw conclusions in Section~\ref{con}.
   
 \section{Methods}
 \label{meth}
 
 The goal of this study is to generate populations of satellite galaxies, including their luminosities and \vmax, for a range of dark matter halo masses and dark matter models, and then compare the results to the measured \vmax of the Milky Way satellites. We first discuss our semi-analytic model of galaxy formation, and then our implementation of the algorithm for calculating the stripping of satellites galaxy haloes. We then present a brief discussion of the observational data, and end with a presentation of the statistic with which we compare the simulated and Milky Way satellite \vmax distributions. We end in Subsection~\ref{SNmps} by expanding our analysis to WDM with a presentation of our sterile neutrino models.
 
 \subsection{Semi-analytic model of galaxy formation}
 
In this Subsection we describe how we generate merger trees for dark matter haloes, and populate the subhaloes with galaxies by means of a semi-analytic model.

 In order to produce populations of satellite galaxies, we generate $5000$ halo merger trees using the algorithm introduced by \citet[][PCH]{Parkinson08}, itself an evolution of the extended Press-Schechter (EPS) algorithm \citep{Bond91} for combinations of a dark matter model and a central halo mass. We have selected 14 host halo masses in the range $[0.5,1.8]\times10^{12}\Msun$, and for most of this paper we focus on three in particular: $0.5\times10^{12}$, $1.0\times10^{12}$, and $1.4\times10^{12}\Msun$. This method is modified for the sterile neutrino models to incorporate a sharp $k$-space filter, as opposed to the standard real space top hat filter, because the latter introduces spurious haloes at small scales \citep{Schneider13, Benson13, Lovell16}. 
 
  The merger trees are then populated with galaxies by means of the {\sc galform} semi-analytic model of galaxy formation \citep{Cole00}. 
  In this study we use a variation of the model described in \citet{Lacey16}, run on dark matter merger trees produced assuming a Planck cosmology \citep{PlanckCP13}. When changing cosmologies, some of the model free parameters needed to be changed in order to still recover a good agreement with the set observations used during its calibration (as described in \citealp{Lacey16}). We refer to this model hereafter as LC16. The features of this model include star formation, supernova feedback, and dynamical friction in the mergers of galaxies.  A full description of the model run assuming an underlying Planck cosmology will be presented in Baugh~et~al.~(in~prep.) Leading semi-analytic models such as this enable us to attach luminosities to the PCH haloes and subhaloes, and thus develop \vmax functions for sets of satellites for which their observations can be reasonably assumed to be complete.
  
  Semi-analytic galaxy formation models vary in their predictions for
  the galaxy population, in particular for satellite galaxies. We
  therefore also
  employ a second version of {\sc galform} as published in
  \citet{GuoQuan16}, (hereafter referred to as G16) to demonstrate the uncertainties
  arising from the galaxy formation model; a full description of this model will be presented in Baugh~et~al.~(in~prep.). This model differs from
  LC16 in two ways that are of interest to this study: the feedback in
  small galaxies is weaker, and the initialisation of orbits is different. In order to show the effects of these two model features,
  we also consider a hybrid model in which the satellite orbits are
  initialised in the same way as in LC16 but all other features and
  parameters are drawn from G16; we label this model as G16-2. Both LC16 and G16 have also been re-calibrated relative to the models published in \citep{Lacey16} and \citet{GuoQuan16} to take account of a satellite merging model developed by \citet{Campbell15} and \citet{Simha16}. However, this merging model is not used here because it requires N-body merger trees as an input. Details will be presented in Baugh~et~al.~(in~prep.) and Gonzalez-Perez~et~al.~(in prep.). A more careful study would ensure that the parameters are re-calibrated self-consistently for the merging model and the cosmological parameters: we differ this work to a future study.
  
  Given the choice of LC16 and G16 for our fiducial model, we select LC16 because it fits a wider range of astronomical observables and in particular gives a better fit to the satellite luminosity function (see Fig.~\ref{LF3}). We consider the impact of the change in models in appendix~\ref{app}. For the remainder of this paper, we use the LC16 model except where explicitly stated otherwise. For all of our models we use the Planck cosmological parameters: $h=0.6777$, $\Omega_{0}=0.304$, $\Omega_{\Lambda}=0.696$, $n_\rmn{s}=0.9611$, $\sigma_{8}=0.8288$ \citep{PlanckCP13}.

 The application of the both versions of {\sc galform} has to be adjusted for the purposes of WDM models. We discuss this issue in Subsection~\ref{SNmps}.

 \subsection{From $V_\rmn{vir}$ at infall to \vmax at $z=0$}
 \label{VvVm}
  
  The PCH algorithm calculates the number of haloes of a given virial mass and virial circular velocity merging onto a host halo at a given redshift, $z_\rmn{infall}$. Two properties that are not predicted by the PCH algorithm are the maximum circular velocity of the object (which is distinct from the virial circular velocity) and the dark matter mass loss of that object. In this subsection we discuss the derivation of these quantities.
  
 We begin with the conversion from virial circular velocities, $V_\rmn{vir}$, to maximum circular velocities, \vmax. These two quantities are related by the equation:
 
 \begin{equation}
 	V_\rmn{max} = 0.465~V_\rmn{vir}\sqrt{\frac{c}{\ln(1+c)-c/(1+c)}},
	\label{eqn:vm}
\end{equation}

\noindent where $c$ is the Navarro-Frenk-White \citep[NFW;][]{NFW_96,NFW_97} profile concentration of the halo as calculated from the halo mass-concentration relation by the {\sc galform} code at the halo formation time.

 One needs to take account of the effects of baryons on the halo mass and \vmax. \citet{Sawala16a} showed that the isolated dwarf haloes experienced a decrease in their \vmax relative to dark matter-only simulations due to the expulsion of gas, an effect not included in the collisionless PCH formalism. They showed that the average magnitude of this suppression, $p=V_\rmn{max,SPH}/V_\rmn{max,DMO}$ takes the following form:
 
\begin{equation}
	p = \begin{cases}    
          0.87 & 0 \le V_\rmn{max,DMO} < 30~\rmn{km~s}^{-1} \\
          g\log_{10}(V_\rmn{max,DMO})+c & 30 \le V_\rmn{max,DMO}< 120~\rm{km~s}^{-1} \\
          1 & 120 \le V_\rmn{max,DMO} \\
  \end{cases}
   \label{eqn:torus}
\end{equation}

  \noindent
  where $g$ and $c$ are the constants required to make the relation continous; a similar relation has been determined for more massive haloes by \citet{Schaller15}. The \vmax and virial mass $m$ are thus adjusted to $V_\rmn{max}=pV_\rmn{max,PCH}$ and $m=p^{2}m_\rmn{PCH}$, where PCH denotes the values output by the PCH algorithm. We present results in which this modification is both present and omitted in order to show the impact of supernova feedback on the fit to the observed \vmax function. We also assume that the concentration of the halo is unaffected by this alteration, and that the stripping procedure developed by \citet{JiangF16} is still an accurate model for the subhalo mass evolution.

  In order to calculate the $z=0$ \vmax functions for our populations of satellites we implement the method of \citet{JiangF16}. The rate of mass loss for the satellite, $\dot m$, at a time $t$ after accretion, is assumed to be given by the equation:

\begin{equation}
	\dot m = A \frac{m(t)}{\tau_\rmn{dyn}}\left(\frac{m(t)}{M(z)}\right)^{\alpha},
	\label{eqn:mdot}
\end{equation}	    

\noindent
where $A$ and $\alpha$\footnote{\citet{JiangF15} denote this parameter as `$\gamma$'. We instead use $\alpha$ in order to avoid confusion with the {\sc galform} feedback power law index.} are parameters to be fitted from N-body simulations, $\tau_\rmn{dyn}$ is the dynamical time, $m(t)$ is the mass of the subhalo at time $t$. $M(z)$ is the mass of the host halo at redshift $z$, and is calculated using the code developed by \citet{Correa15a,Correa15b,Correa15c}. \citet{JiangF15} and \citet{JiangF16} fit $\alpha=0.07$, and a mean of  $A$, $\bar A=0.86$. They extract sample values of $A$ from a log-normal distribution using this $\bar A$ and a standard deviation of 0.17. however,  we recalibrate this parameter for our work (see below).

The dynamical time is calculated based on the estimated overdensity of the halo at each redshift, denoted $\Delta_{c}$. The relationship between the two is:

\begin{equation}
	\tau_\rmn{dyn}=\frac{1.628/h}{\sqrt{\Omega_{0}(z+1)^3+\Omega_{\Lambda}}  }\left(\frac{\Delta_{c}}{178}\right)^{-0.5},
\end{equation}

\noindent and $\Delta_{c}$ itself is given by:

\begin{equation}
	\Delta_c = 18\pi^2 + 82(\Omega(z)-1)-39(\Omega(z)-1)^2,
\end{equation}

\noindent
where $\Omega(z)$ is value of the cosmological matter density parameter at redshift $z$, as shown by \citet{Bryan98}. 

The next step is to translate the change in virial mass into a change in \vmax,  which is achieved via the relation fitted to simulations in \citet{Penarrubia08b} and \citet{Penarrubia10}:

\begin{equation}
	V_\rmn{max}(z=0)=1.32~V_\rmn{max}(z=z_\rmn{infall})\frac{x^{0.3}}{(1+x)^{0.4}},
	\label{eqn:VmVv}
\end{equation} 

\noindent where $x$ is the ratio of the redshift zero mass to the infall mass, i.e. $x=m(z=0)/m(z=z_\rmn{infall})$. 

\begin{figure}
     \includegraphics[scale=0.5]{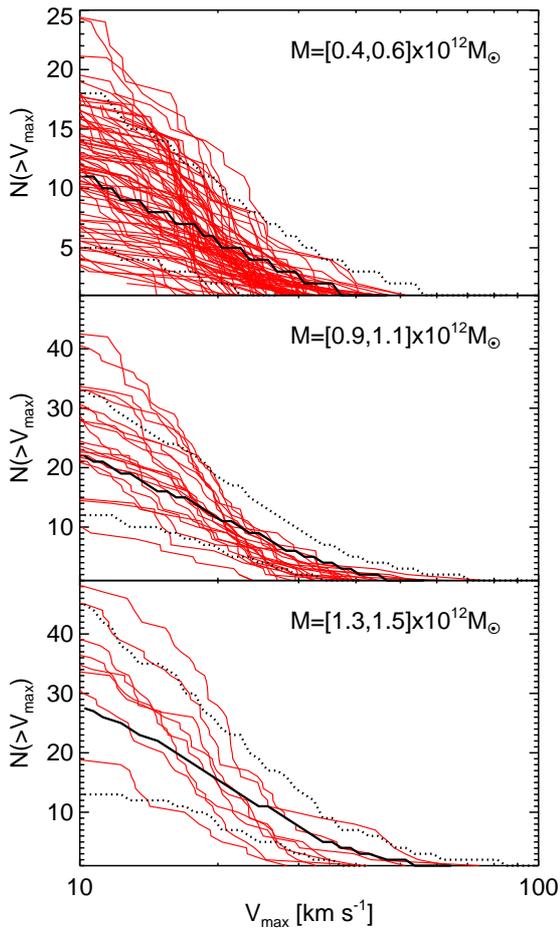}
     \caption{\vmax functions in CDM N-body simulations and those computed using the PCH+stripping method described in Subsection~\ref{VvVm}. The COCO \vmax functions are shown as red lines, and for the PCH functions we show the median \vmax function across $\sim1000$ haloes as a solid black line and the 68~per~cent scatter regions by dotted lines. In both cases we restrict the selection to include only subhaloes for which the peak value of \vmax, $V_\rmn{peak}>20$~\kms. The three panels show  results for three central halo mass bins: $[0.4,0.6]\times10^{12}\Msun$ (top panel), $[0.9,1.1]\times10^{12}\Msun$ (middle), and $[1.3,1.5]\times10^{12}\Msun$ (bottom). The distribution of PCH halo masses across each bin is determined according to the halo mass function.}
     \label{VmCOCO}
 \end{figure}

As a check of our method, we compare the results of our computation to
those of N-body simulations. In Fig.~~\ref{VmCOCO} we plot the \vmax
functions for subhaloes in the CDM-COCO simulation \citep{Hellwing16}, a zoomed N-body simulation with a high resolution region of radius $\sim17h^{-1}\rmn{Mpc}$ and simulation particle mass of $1.1\times10^5\Msun$.
Here we include subhaloes out to the radius of spherical top-hat
collapse, $r_\rmn{th}$, in order to be consistent with the PCH
algorithm outputs. We also plot the median of $\sim100-700$ (highest mass-lowest mass bin) \vmax functions in
which we retain all subhaloes that had an accretion \vmax greater than
$20$~\kms irrespective of whether they host a satellite galaxy, with
the exception of those subhaloes that are located within other
subhaloes at redshift zero. For this comparison, we also do not apply the \citet{Sawala16a}
correction since COCO is a dark matter-only simulation. We select COCO
haloes in the following mass brackets:
$[0.4,0.6]\times10^{12}\Msun$,$[0.9,1.1]\times10^{12}\Msun$, and
$[1.3,1.5]\times10^{12}\Msun$, and the masses we use are the mass
enclosed within $r_\rmn{th}$. The PCH masses are drawn from the same
brackets in mass, and are selected to fit the halo mass function of
\citet{Jenkins01}. In both the N-body subhalo and semi-analytic galaxy cases we select only objects that are substructures of the host halo rather than substructures of satellites. 

There is good agreement between the semi-analytic model and the simulation at \vmax$>20$~\kms. In order to achieve this agreement we recalibrated the $\bar A$ parameter to $\bar A=1.4$. The semi-analytic model lacks the satellites with  \vmax$>80$~\kms, and slightly overpredicts the upward scatter around  \vmax$\sim24$~\kms in the middle mass bin. The model consistently underpredicts the number of satellites with  \vmax$<20$~\kms. Potential causes of this discrepancy include the presence of subsubstructure in the simulation data and a tendency for the model to over-strip small haloes. One should also bear in mind that the COCO simulation was performed with the {\it WMAP-7} cosmological parameters rather than {\it Planck}, therefore a careful study will require that the model be calibrated against simulations using {\it Planck}.

\subsection{Milky Way satellite properties}

Our analysis requires two properties of the Milky Way satellites, their $V$-band magnitudes and their \vmax values. We source our $V$-band values from the data set compiled by \citet{McConnachie12}, and measured by \citet{deVaucouleurs91,Irwin95,Martin08}. The \vmax are much more difficult to measure, and typically involve some cross-correlation with CDM simulations. One example of this is the work of \citet{Sawala16a}, who use high resolution hydrodynamical simulations to derive likely \vmax values for nine of the dwarf spheroidals based on the simulated satellite's luminosities and central densities. This method has the advantage of selecting the haloes that are most likely to host satellites, whose \vmax is biased relative to CDM simulation expectations. We therefore use \vmax plus associated error bars derived from \citet{Sawala16a} where available. For satellites not included in their study we use the \vmax values and error bars collated in \citet{JiangF15}, which were obtained using a likelihood analysis of the satellite velocity-dispersion \citep{Kuhlen10,BoylanKolchin12}, and rotation curves \citep{vanderMarel14}.

\subsection{Likelihood from \vmax distributions}
\label{sub:SL}

Here we summarise the statistical method for comparing the observational and simulated \vmax distributions. It is identical to that of \citet{JiangF15} except where noted below.

The goal is to determine the probability that the Milky Way satellite \vmax function can be drawn from the distribution of simulated functions. We will establish the statistical scatter between the simulated \vmax functions, and calculate the mean deviation between the measured Milky Way \vmax function and the simulated systems. The size of the measured \vmax function deviation relative to the size of the scatter will tell us about how readily the Milky Way \vmax function is realised in each of our models.

The first step is to define the variation within the set of \vmax for a given halo mass-dark matter model-galaxy formation model combination. The $n$ most massive satellites of the $i$-th simulated halo are selected, and their values of \vmax are sorted into descending order. The \vmax distribution is then \{$V_{i,1},V_{i,2},V_{i,3} ...V_{i,n}$\}; here we have omitted the `max' subscript for clarity. We can define the difference between this $i$-th halo distribution with respect to the $j$-th halo distribution thus:

\begin{equation}
	Q_{i,j} = \frac{\sum_{k=1}^n|V_{i,k}-V_{j,k}|}{\sum_{k=1}^n(V_{i,k}+V_{j,k})},
	\label{qij}
\end{equation}

\noindent
and if there are $N$ realisations of the model in question, the mean $Q$ for the $i$-th distribution, $\bar Q_{i}$, is:

\begin{equation}
	\bar Q_{i} = \frac{1}{N-1}\sum_{j\ne i}Q_{i,j}.
\end{equation}

\noindent

Similarly, if we substitute the $i$-th simulated \vmax distribution to instead be \{$V_{\rmn{MW},1},V_{\rmn{MW},2},V_{\rmn{MW},3} ...V_{\rmn{MW},n}$\}, i.e. the observed \vmax distribution of the Milky Way satellites, then we obtain $Q_\rmn{MW}$:

\begin{equation}
	 Q_\rmn{MW} = \frac{1}{N}\sum_{j}Q_\rmn{MW,j}.
\end{equation}
 
 \noindent
 
 The probability that a \vmax function with $Q_\rmn{MW}$ could be drawn from the parent distribution is then $P(>Q_{MW})$, where $P$ is the cumulative distribution of $\bar Q$.

 We expand on the method described above to describe how we select
 satellites. The luminosities calculated for the satellites enable us
 to take account of the fact that the brightest satellites, for which
 the velocity dispersions have been measured with the highest
 precision, need not necessarily reside in the most massive haloes.
 Therefore, we consider two options for selecting our top `$n$'
 satellites to be matched to observations: \emph{i)} select the $n$
 brightest $V$-band satellites, and \emph{ii)} select the the $n$
 highest \vmax satellites. We compare the results from these two
 approaches in Subsection~\ref{sub:statcomp}.

 \subsection{Sterile neutrino matter power spectra}
 \label{SNmps}

In addition to CDM, we consider keV-scale, resonantly-produced sterile neutrino dark matter. The latter constitutes part of a larger particle physics model called the neutrino minimal standard model ($\nu$MSM), which explains neutrino oscillations and baryogenesis in addition to yielding a dark matter candidate, see \citet{Boyarsky09a} for a review. The keV sterile neutrino behaves like WDM, in that it free-streams out of small perturbations in the early Universe.  The resulting matter power spectrum cutoff is influenced by two parameters: the sterile neutrino mass, $M_\rmn{s}$, and the lepton asymmetry in which the dark matter is generated \citep{Shi99, Boyarsky09a, Laine08, Ghiglieri15, Venumadhav15}. We parametrize the lepton asymmetry as $L_6$, which is defined as $10^{6}\times$ the difference in lepton and anti-lepton abundance normalised by the entropy density. The power spectrum cutoff shifts to smaller scales for larger values of the mass, as is the case for thermal relic WDM. By contrast, the behaviour with lepton asymmetry is non-monotonic; for a recent discussion see \citet{Lovell16}. 

We focus on the parameter space that is roughly in agreement with the recent observations of the 3.5~keV emission line detected in \citet{Bulbul14,Boyarsky14a,Boyarsky15}, which requires a sterile neutrino mass of 7~keV and a lepton asymmetry in the range $L_6=[8,11.2]$, where the uncertainty in $L_6$ is dominated by the uncertainty in the dark matter content of the target galaxies and galaxy clusters. The recent study by \citet{Ruchayskiy15} set a more stringent lower limit of $L_6>9$; however, $L_6=8$ remains of interest as it has the shortest free-streaming length obtainable by a 7~keV sterile neutrino of any lepton asymmetry. We therefore select primarily three models for our study, $L_6=8,10,12$, in order to span the range of $L_6$ that is in agreement with the detected decay line. From hereon in we refer to these models as LA8, LA10, and LA12. We also briefly consider four further models to probe a larger range of free streaming lengths: three 7~keV particles ($L_6$=[14,18, 120]) and one 10~keV sterile neutrino with $L_6=7$.

\begin{figure}
     \includegraphics[scale=0.33]{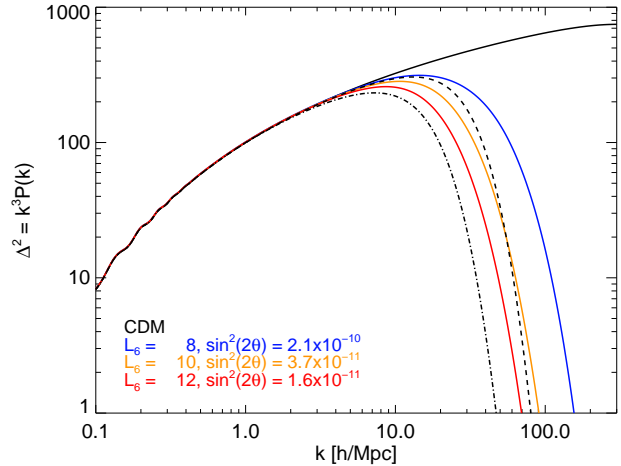}
     \caption{Matter power spectra for our four dark matter models: CDM (black, solid), LA8 (blue), LA10 (orange) and LA12 (red). The black dot-dashed line denotes the power spectrum of the 2.3~keV thermal relic studied in \citet{WangMY15}, and the dashed line is the 3.3~keV thermal relic power spectrum used in COCO-WDM \citep{Bose16a}.}
     \label{PS}
 \end{figure}

We first calculate the momentum distribution functions for our three sterile neutrino models using the methods and code of \citet{Laine08} and \citet{Ghiglieri15}. From these distribution functions we then derive the matter power spectra by means of a modified version of the {\sc camb} Boltzmann-solver code \citep{Lewis2000}.  The results are plotted in Fig.~\ref{PS} as dimensionless matter power spectra. All three models exhibit a cutoff, and the cutoff position shifts to larger scales -- smaller wavenumbers -- with increasing $L_6$\footnote{$L_6=8$ is the model for which the cutoff is located at the smallest scale, since for smaller $L_6$ the influence of resonant production is weaker and thus the cutoff moves to larger scales.}. 

Also plotted is the power spectrum of the 2.3~keV thermal relic
studied by \citet{WangMY15}, who showed, using N-body simulations,
that, since halo concentrations are lower for WDM than for CDM haloes,
this particular model required subhaloes of $V_\rmn{max}\sim1.17$
times higher than $\Lambda$CDM to fit the kinematics and photometry of
Fornax. We will use this correction factor in our study to illustrate
the impact of lower sterile neutrino halo densities on their hosted
galaxies. We caution that this factor was derived for only one
satellite and for a dark matter model that has a larger free-streaming
length than any of our three primary WDM models. Our results for WDM
should therefore be considered as a rough approximation, rather than
rigorous predictions. In addition, central halo masses
$<1.4\times10^{12}\Msun$ are disfavoured for these models in the
current model of reionization feedback by virtue of their low
satellite counts \citep{Lovell16}; however, we include them here for
completeness.

 The application of the {\sc galform} feedback model is complicated in WDM-style models by the dependence of the feedback strength on the halo circular velocity. In {\sc galform}, the strength of feedback is modelled as a power law of the circular velocity, where the power law index is denoted $\gamma$. The lower circular velocities of WDM haloes lead to the result that WDM models run using the CDM model parameters under-predict the number of galaxies with $M_V<-16$.  A discussion of this issue can be found in \citet{Kennedy14} and \citet{Lovell16}. We recalibrate the model against the $b_{J}$ band luminosity function and find that $\gamma=3.15$ is a good fit to the observational data for all three of our sterile neutrino models as opposed to $\gamma=3.4$ for the standard CDM model. We therefore adopt $\gamma=3.15$ for LA8, LA10, and LA12, and retain $\gamma=3.4$ for CDM.
 
 %%%%%%%%%%
 
 We also make the following assumptions with regard to the stripping algorithm in the sterile neutrino models:
 
 \begin{itemize}
\item Given that WDM subhaloes deviate slightly from NFW profiles \citep{Colin08, Lovell14,Ludlow16}, a complete study would re-evaluate whether the \vmax-$V_\rmn{vir}$ relation (equation~\ref{eqn:VmVv}) would need to be recalibrated. For simplicity we use equation~\ref{eqn:VmVv} to calculate \vmax for all of our models.

\item Hydrodynamical models of WDM have shown that WDM subhaloes exhibit the same degree of mass loss as CDM haloes \citep{Lovell17}, thus equation~\ref{eqn:torus} is equally valid for our sterile neutrino simulations.

%\item A self-consistent analysis would recalibrate the stripping normalization parameter, $\bar A$, for each of our sterile neutrino models. However, we do not have access to any simulations with a large number of haloes that use the matter power spectra, therefore we assume that $\bar A=1.4$ works equally well for WDM models. In practice, we are likely overestimating the stripping of WDM haloes since WDM concentrations are lower yet we are assuming CDM concentrations. Our results are therefore more extreme than the recalibrated \vmax functions would be. \footnote{A WDM version of the COCO volume has been run \citep{Bose16a}; however, the free-streaming length used was shorter than any of our sterile neutrino models and therefore not suitable for a precision calibration}. 
\end{itemize}

\begin{figure}
     \includegraphics[scale=0.5]{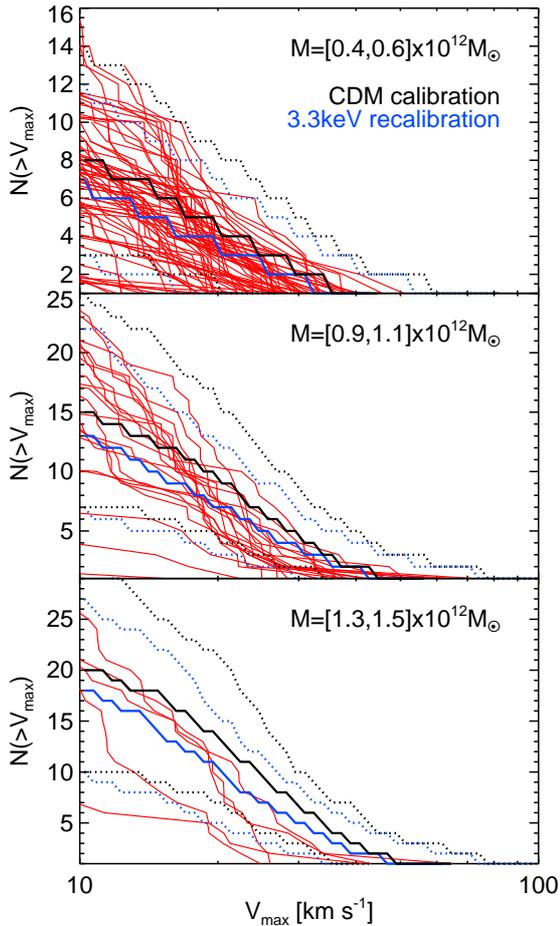}
     \caption{The \vmax functions of 3.3~keV thermal relics as predicted by the COCO-WDM simulation and the PCH+stripping method. We include the PCH data as computed using the CDM calibration ($\bar A=1.4$) in black and the re-calibration for the 3.3~keV relic ($\bar A=1.1$) in blue. The COCO-WDM \vmax functions are shown as red lines.}
     \label{VmCOCOw}
 \end{figure}

 The stripping model is calibrated to CDM simulations,  in which the halo mass-concentration relation will play a key role in the stripping rates. This relation changes for, and between, different WDM models. Therefore, a precise prediction for the $z=0$ \vmax functions for a given WDM model requires that we calibrate each model to an N-body simulation of that specific model. We do not have N-body simulations for any of the sterile neutrino models discussed below; instead we make a qualitative prediction for how our results would change by calibrating our model to that of a 3.3~keV relic as used in the COCO-WDM simulation \citep{Bose16a}, which is a good approximation to our LA8 model. We repeat the same process discussed above as applied to COCO-CDM, with the CDM matter power spectrum replaced by that of a 3.3~keV thermal relic, both with the CDM-calibration value $\bar A=1.4$ and a re-calibrated version with $\bar A=1.1$. We present our results in Fig.~\ref{VmCOCOw}.

The original calibration works well for the lowest mass halo bin,  but systematically overpredicts the \vmax functions of the $1.0\times10^{12}\Msun$ and $1.4\times10^{12}\Msun$. This is because the WDM haloes are less concentrated than the CDM and thus the stripping rates are higher. Our recalibration ameliorates some of the discrepancy, a lthough it still overpredicts the \vmax functions of the two more massive haloes, in order to not {\it under}-predict the $0.5\times10^{12}\Msun$ mass functions. The mean suppression of the recalibrated model relative to the original at a \vmax of 20\kms of 30~per~cent, even for this relatively warm model, and is therefore significant. We adopt $\bar A=1.1$ for all of our sterile neutrino models, and state how the results would change for a precise calibration to each separate model where appropriate.

 \section{Results}  
  \label{res}
  
  In this section we show to what degree our models agree with the observed luminosities and \vmax of the Milky Way satellites, and then analyse the Too Big To Fail problem using the  statistic developed by \citet{JiangF15}. We first consider the effect of baryon physics on the CDM \vmax function in Subsection~\ref{ss:bp}, and then apply our preferred baryon model to the sterile neutrino models in Subsection~\ref{ss:sndm}.
  
  \subsection{Baryon physics}
  \label{ss:bp}
  
  \subsubsection{Luminosity functions}
  
\begin{figure}
     \includegraphics[scale=0.5]{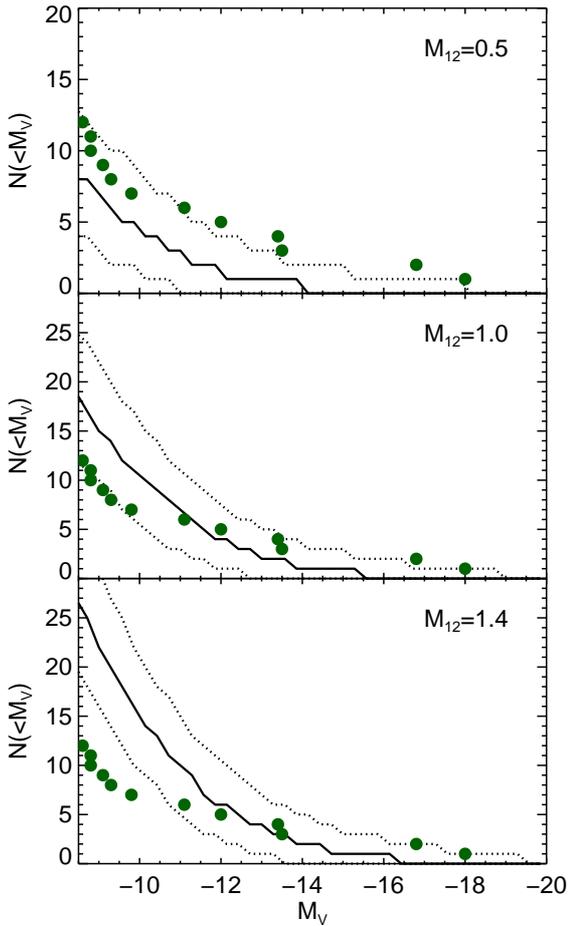}
     \caption{Cumulative satellite luminosity function for the LC16 $\Lambda$CDM galaxy
       formation model and three halo masses. The solid lines denote
       the median number of satellites brighter than $M_{V}$ across
       all the realisations and the dotted curves mark the 5 and
       95 percentiles. The top, middle, and bottom panels show the
       mass functions for central haloes of mass 0.5, 1.0, and
       $1.4\times10^{12}\Msun$ respectively. The circles mark the
       observed Milky Way satellite luminosity function.}
     \label{LF}
 \end{figure}

  We begin our discussion of the results with the luminosity functions of each of our {\sc galform} models. In Fig.~\ref{LF} we present the luminosity functions for the LC16 model and three halo masses ($[0.5,1.0,1.4]\times10^{12}\Msun$). We also include the observed luminosity function of the Milky Way satellites, which we assume to be complete for the range of luminosities considered. 
  
  The most striking difference between the observations and all four models is the steepness of all the simulated luminosity functions relative to that of the Milky Way. This is realised as a dearth of large and small Magellanic cloud (LMC and SMC) candidates for the  $5\times10^{11}\Msun$ halo and an overproduction of bright satellites for the central mass of $1.4\times10^{12}\Msun$. However, the $1\times10^{12}\Msun$ returns a reasonable match to the observations.
    
 \subsubsection{\vmax functions for luminous satellites}
 
\begin{figure}
     \includegraphics[scale=0.5]{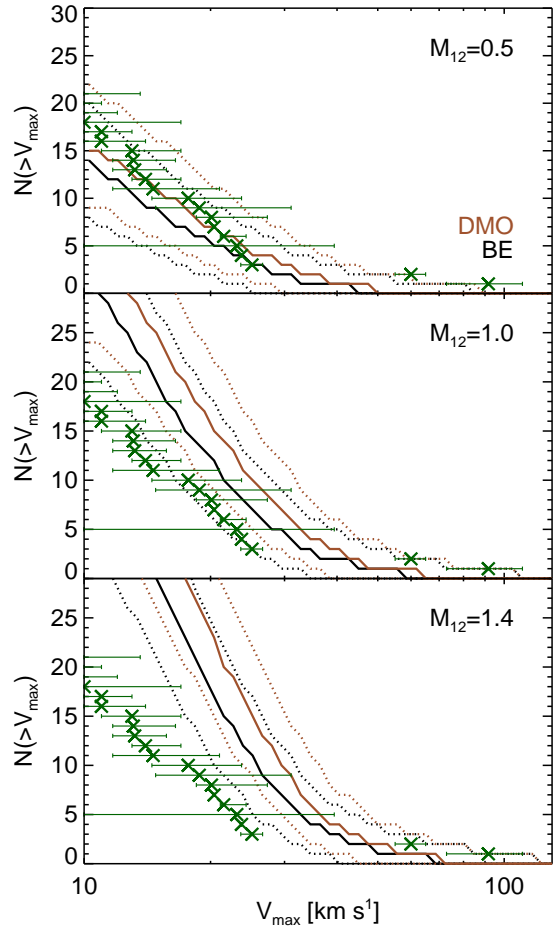}
     \caption{Cumulative satellite \vmax function for the
       $\Lambda$CDM-LC16 model when the correction for baryonic effects 
     is applied (black) and not (brown). The top, middle,
       and bottom panels show the mass functions for central haloes of
       mass 0.5, 1.0, and $1.4\times10^{12}\Msun$ respectively. The
       inferred \vmax function of the Milky Way satellites is shown as
       green crosses. Note that error bars are not included for two of
       the satellites, Fornax and Draco, because these were not
       calculated by \citet{Sawala16a}.}
     \label{VF}
 \end{figure}

 Identifying which satellites are luminous enables us to make a more
 accurate comparison between the simulated \vmax function and that
 inferred for the Milky Way satellites. The \vmax function is
 influenced by early loss of baryons from a halo, as described by \citet{Sawala16a}. We
 illustrate the importance of this effect in Fig.~\ref{VF}, in which
 we show the median cumulative \vmax functions for CDM in two cases,
 with the baryon suppression of equation~\ref{eqn:torus} turned off
 (brown lines) and turned on (black lines). Unlike in Fig.~\ref{VmCOCO}, we
 only plot satellites that are luminous.

%. The amplitude of the functions increases with central mass and decreases with free-streaming length, much as was the case for the luminosity functions. The tension between CDM and the measured \vmax function is apparent even for $V_\rmn{max}\sim30~\rmn{kms}^{-1}$. In addition to the \vmax function derived from CDM simulations, we also show a \vmax function modified to take account of the fact that WDM haloes host less dense galaxies. The agreement is worse for the lowest halo mass, but for $M>1\times10^{12}\Msun$ the fit to the sterile neutrino models is good, but for the lack of LMC and SMC analogues. However, note that this correction is derived for just one satellite (Fornax) and one WDM model (2.3~keV thermal relic), therefore this result should be considered illustrative rather than a quantitive prediction.

 We first discuss the case in which the suppression of \vmax by
 baryon effects is not taken into
 account. For the lowest mass halo, the $\Lambda$CDM model provides a
 good description of the data, except for the lack of any LMC
 counterparts. For a halo mass of
 $1.0\times10^{12}\Msun$ the model tends to overpredict the
 observed \vmax function, although the uncertainties in \vmax are
 large enough for the model to be consistent with the data. For a halo
 mass of 
 $1.4\times10^{12}\Msun$ the discrepancy is large enough that it 
 cannot be explained by uncertainties in the \vmax
 measurements. Applying the correction to \vmax due to
 baryon effects, as described by
 equation~\ref{eqn:torus}, produces a  significant shift in the
 predicted \vmax function. Now the  models with halo masses of
 $0.5$ and $1\times10^{12}\Msun$  are entirely
 consistent with the data but the model with the largest halo mass is
 still ruled out. We therefore conclude that the
 suppression of satellite mass caused by the early loss of baryons
 from the halo is crucial in order to explain the observed \vmax
 function, in agreement with \citet{Sawala16a} and \citet{Fattahi16},
 but with the added constraint that the mass of the Milky Way halo
 should be lower that about $1.4\times10^{12}\Msun$.
 
 \subsubsection{Statistical comparison of simulated and observed \vmax functions}
 \label{sub:statcomp}
 
The strength of our PCH method, as compared to hydrodynamical simulations like \citet{Sawala16a} and \citet{Fattahi16}, is that it is practical to run hundreds of merger trees very quickly and thus build good statistical samples. We can therefore calculate what proportion of simulated systems returns a \vmax function that is a good match to that of the Milky Way satellites, and thus quantify the quality of the agreement between observations and the model \vmax functions shown in Fig.~\ref{VF}. This is done by extracting the $n$ most massive luminous satellites and calculating the $Q$ statistic for this distribution using the methods of \citet{JiangF15} as summarised in Subsection~\ref{sub:SL}. We compare the value of $Q$ obtained for the Milky Way system with respect to the PCH results, denoted $Q_\rmn{MW}$, to the distribution of PCH $Q$. The closer $Q_\rmn{MW}$ is to the centre of the $Q$ distribution, the better the agreement is between the model and the observations.

\begin{figure}
     \includegraphics[scale=0.5]{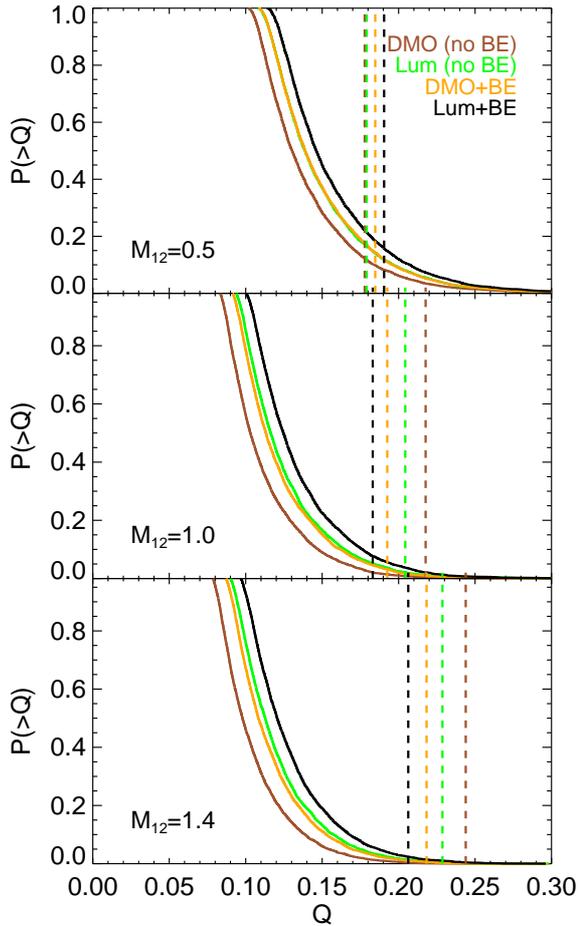}
     \caption{Cumulative  $Q$ statistic function for the CDM-LC16 model using our four \vmax variations: luminous satellites (brown), baronic effects applied (orange), luminosity ordered satellites (green), and luminosity ordered with baryonic effects applied (black).  Solid lines denote the cumulative $Q$ statistic functions, and dashed lines the corresponding value of $Q_\rmn{MW}$. The top, middle, and bottom panels show the mass functions for central haloes of mass 0.5, 1.0, and $1.4\times10^{12}\Msun$ respectively. }
     \label{QF}
 \end{figure}

In Fig.~\ref{QF} we plot the distributions of $Q$ for the $0.5\times10^{12}$, $1.0\times10^{12}$, and $1.4\times10^{12}\Msun$ haloes and four algorithms for generating \vmax functions. These algorithms are:

\begin{enumerate}
	\item All satellites, baryon effects not applied (also referred to as `DMO').
	\item All satellites, baryon effects included (BE).
	\item Satellites ordered by luminosity, baryon effects not applied (Lum).
	\item Satellites ordered by luminosity, baryon effects included (Lum+BE).
\end{enumerate}

 The number of satellites selected in each case is $n=10$.  
   
For all three halo masses we measure an important effect on the $Q$ distribution between the different algorithms. The application of the feedback suppression factor increases the scatter slightly between distributions relative to the base model (model (i) above) and thus translates each curve to the right by 0.01 units in $Q$ irrespective of the halo mass. A marginally larger shift occurs when satellites are first sorted by luminosity, and the two effects combined produce a shift of 0.02 $Q$ relative to the base. 

There is also a trend on the value of $Q_\rmn{MW}$. When considering the haloes of mass $1.0\times10^{12}\Msun$ and $1.4\times10^{12}\Msun$, luminosity ordering lowers $Q_\rmn{MW}$ as the greater scatter grows closer to encompassing the observational data. The baryonic effects produce a stronger effect in the same direction because the increase in the {\it scatter} is accompanied by a fall in the {\it mean} \vmax function,  and thus closer to the Milky Way satellite \vmax function as shown in Fig.~\ref{VF}. The application of these two lower $Q_\rmn{MW}$ still further, by a total of 0.04 points relative to the base model.  In combination with the greater scatter within the simulated distributions, the overlap between $Q_\rmn{MW}$ and the $Q$ distributions improves significantly. Halo masses that would have been incompatible with the Milky Way satellite \vmax function under the base model are now very possible, if still rare. Note that this improvement does not occur for the lightest halo mass;  however, the base model \vmax function is itself in good agreement with that of the Milky Way satellites, so further suppression results in stronger disagreement with the data. 
 
 \subsubsection{Probability of drawing the Milky Way satellite \vmax function from simulated \vmax distributions}
 \label{sss:pq}
 
\begin{figure}
     \includegraphics[scale=0.33]{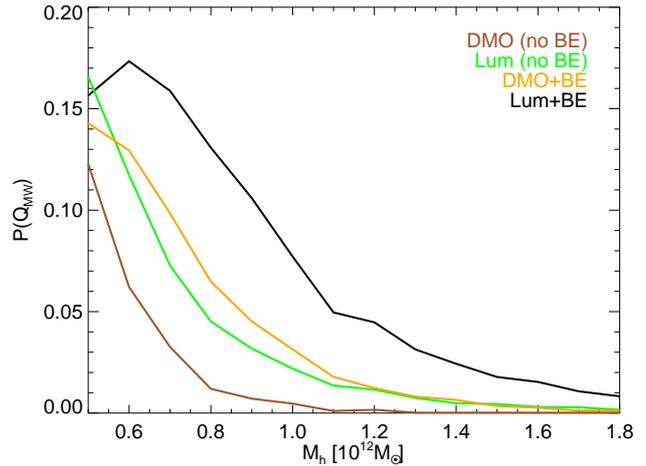}
     \caption{The probability that a Milky Way-like satellite \vmax
       distribution is drawn from the simulated distributions as a
       function of halo mass. Our varieties of \vmax functions are
       shown using the same colours as in Fig.~\ref{QF}: luminous
       satellites (brown), luminosity ordered satellites (green),
       baryonic effects applied (orange), and luminosity ordered with
       baryonic effects applied (black).}
     \label{PP}
 \end{figure}

In the previous subsubsection we showed that the mean \vmax function amplitude correlates with central halo mass, such that for a given halo selection algorithm there is a `sweet spot' halo mass at which the probability of drawing a Milky Way-like satellite \vmax function is maximised. The probability that the Milky Way distribution can be drawn from a \vmax distribution at fixed host halo mass is quantified by the cumulative probability distribution $P(>Q_{MW})$. If $P(>Q_\rmn{MW})\ll0.01$ then that halo mass-model combination is ruled out. Therefore, we calculate $P(>Q_\rmn{MW})$ as a function of halo mass for our set of 14 central halo masses and plot the results for our four \vmax function algorithms in Fig.~\ref{PP}. Note that in all four cases we select ten satellites, the difference is solely in how they are selected and processed.  
 
All four curves show preferences for lighter haloes; however, the luminosity-ordered+feedback suppression shows a shift towards higher mass haloes. The amplitude of the curves is lowest for the base model, which registers an effective zero probability for  haloes more massive than $1.3\times10^{12}\Msun$. Luminosity ordering increases the probability across all halo masses, feedback suppression further still, and the highest probabilities are found for the luminosity-ordering+feedback suppression algorithm. In this case, even halo masses of $1.8\times10^{12}\Msun$ can host Milky Way-like satellite \vmax functions, albeit very rarely. One should also note that we showed in Fig.~\ref{VmCOCO} that our stripping model overpredicts the number of $\sim25$\kms subhaloes; therefore, a more accurate stripping model will return probabilities higher than those calculated here.
   
   \subsection{Sterile neutrino dark matter}
  \label{ss:sndm}
  
\begin{figure}
     \includegraphics[scale=0.5]{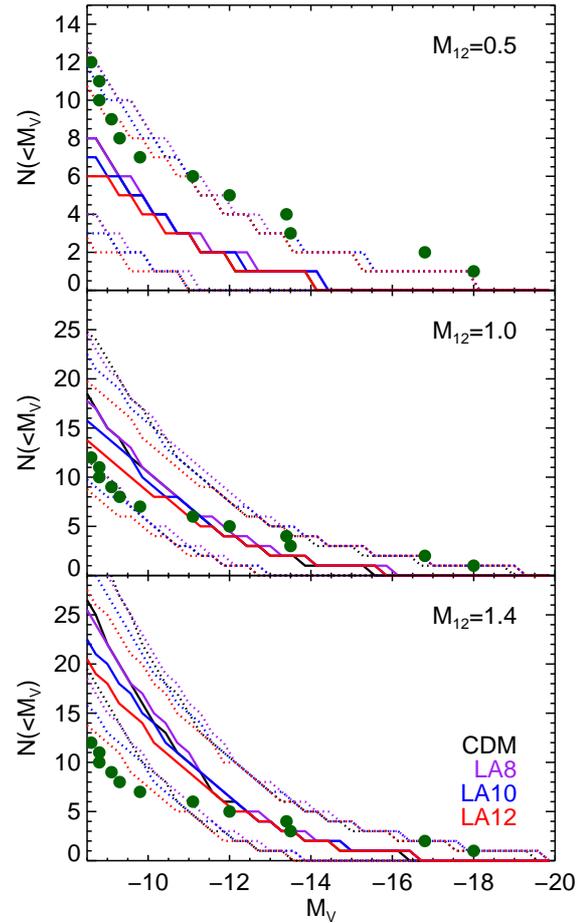}
     \caption{Cumulative satellite luminosity function for four dark
       matter models and three halo masses. The galaxy formation model
       is LC16, with a refitted $\gamma_\rmn{SN}$ parameter for the
       sterile neutrino models. The solid lines denote the median
       number of satellites brighter than $M_{V}$ across all of our
       realisations, and the dotted curves mark the 5 and 95
       percentiles. Each dark matter model is denoted by a different
       colour: CDM (black), LA8 (purple), LA10 (blue) and LA12 (red), as indicated in the legend. The top, middle, and bottom panels show the mass functions for central haloes of mass 0.5, 1.0, and $1.4\times10^{12}\Msun$ respectively. The circles mark the observed Milky Way satellite luminosity function.}
     \label{LFW}
 \end{figure}

  We now consider the changes that would be made to our results if the dark matter were a WDM candidate, specifically the sterile neutrino. In Fig.~\ref{LFW} we plot the luminosity functions of three sterile neutrino models, LA8, LA10, and LA12, in addition to CDM. The luminosity functions between CDM and LA8 are remarkably similar, which is in part due to our use of weaker, re-calibrated feedback. The number of satellites is suppressed in the other two models; however, not enough to achieve agreement with the the observations for the highest mass halo. Any comprehensive and accurate model of galaxy formation would therefore still require stronger feedback in low mass galaxies than that used here, although the adoption of WDM may play a subdominant part in achieving the necessary agreement.
  
\begin{figure}
     \includegraphics[scale=0.5]{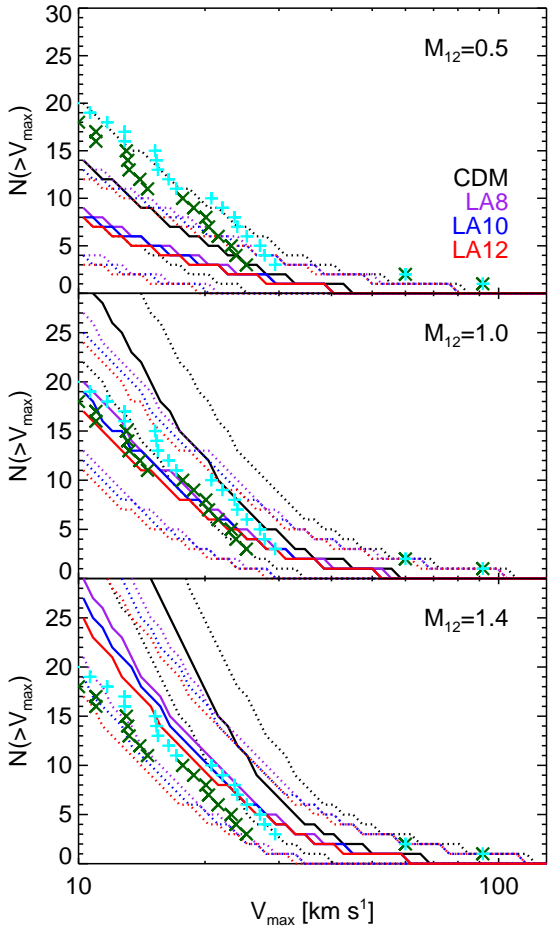}
     \caption{Cumulative satellite \vmax function for the same dark
       matter models and halo masses presented in Fig.~\ref{LFW}. We
       include all luminous satellites, and have applied the baryonic
       feedback correction from equation~\ref{eqn:torus}. The solid
       lines denote the median \vmax across all of our realisations,
       and the dotted curves again mark the 5 and 95 percentiles. The
       colour-dark matter model correspondence is the same as in
       Fig.~\ref{LFW}: CDM (black), LA8 (purple), LA10 (blue), and
       LA12 (red). The top, middle, and bottom panels again show the
       mass functions for central haloes of mass 0.5, 1.0, and
       $1.4\times10^{12}\Msun$ respectively. The dark green crosses
       mark the inferred Milky Way satellite \vmax function assuming
       CDM. We also include cyan plus signs, for which the \vmax
       values of the dwarf spheroidals (but not the Magellanic Clouds)
       are multiplied by the factor of 1.17 suggested by the results
       of \citet{WangMY15}. We have not attempted to correct for
       incompleteness in the observed satellite sample. Therefore
       these values constitute a lower bound on the expected Milky Way
       satellite \vmax function. The \vmax error bars have been
       omitted for clarity.}
     \label{VFW}
 \end{figure}

 Having shown that the sterile neutrino models produce acceptable
 numbers of satellites, we now consider their \vmax functions. We
 apply the suppression factor from baryon effects from
 equation~\ref{eqn:torus} to
 all four dark matter models and plot the results in Fig.~\ref{VFW}.
 There is a systematic decrease of the \vmax function with
 free-streaming length, to the extent that LA12 hosts $\sim50$~per~cent
 fewer satellites with $V_\rmn{max}>10~$\kms than $\Lambda$CDM. This
 suppression moves the sterile neutrino \vmax functions closer to the
 measured Milky Way satellite \vmax function. The improvement is even
 greater for the $1.4\times10^{12}\Msun$ halo when we take into account the different concentration-mass
 relation of WDM models, as parametrised by our dwarf spheroidal \vmax
 correction value of 1.17; for the $1.0\times10^{12}\Msun$ and $0.5\times10^{12}\Msun$ haloes the agreement with the modified \vmax function is instead weaker, since the theoretical \vmax functions are now {\it over}-suppressed. Thus, in general the suppressed \vmax functions and
 lower concentrations of the sterile neutrino models combine to give
 better agreement with the observations at larger halo masses than in
 $\Lambda$CDM.
 
\begin{figure}
     \includegraphics[scale=0.33]{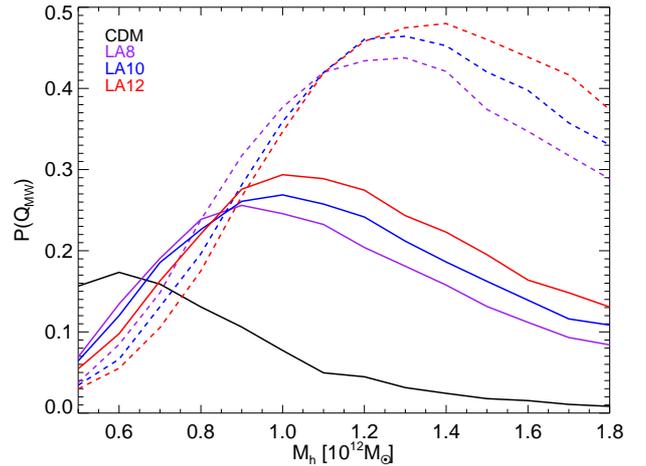}
 
     \caption{The probability that the a Milky Way-like satellite \vmax distribution is drawn from the simulated distributions as a function of halo mass. The \vmax function selection is made using the luminosity-ordered + baryonic effects correction scheme, and the galaxy formation model is LC16 with recalibration for the sterile neutrino models. The colour-dark matter model correspondence is the same as in Fig.~\ref{LFW}: CDM (black), LA8 (purple), LA10 (blue), and LA12 (red). Solid lines denote results calculated when the observed values of \vmax are derived from CDM simulations, and dashed where the \citet{WangMY15} factor is applied. }
     \label{PPW}
 \end{figure}

 To end this subsection, we calculate the probability of drawing Milky Way-like satellite \vmax functions from our sterile neutrino \vmax distributions, once again using the $Q$ distribution-$Q_\rmn{MW}$ combination from Sububsection~\ref{sss:pq}. We present our results as a function of host halo mass in Fig.~\ref{PPW}. When we assume the same values of \vmax for the Milky Way satellites in the sterile neutrino models as in CDM, we find that the amplitude of the probability curve remains roughly the same. The difference instead comes from a shift to larger masses of the probability distribution peak, which reflects how the decrease in the number of satellites requires a more massive host halo to hit the observed target. The consequences at the largest halo masses are significant: a fit to the $1.0\times10^{12}\Msun$ halo is over three times as likely in LA12 than it is in CDM, and  a fit to the $1.4\times10^{12}\Msun$ halo eight times more likely . 
 
 More impressive still is the contribution made by the lower
 concentrations. The adoption of the \citet{WangMY15} correction
 factor improves the probability by over a factor of two as compared
 to the CDM-\vmax values, with the peak in the probability
 distribution located as high as $1.4\times10^{12}\Msun$. This result
 reflects the fact that the observed \vmax function has not only a
 higher amplitude in WDM, which can be achieved just by choosing a
 larger halo, but is also steeper, and therefore has a shape more in
 keeping with that of the simulated data. We stress, however, that
 this result is purely illustrative because it is based on just one
 WDM model (a 2.2~keV thermal relic) and one observed satellite
 (Fornax), therefore fits of many more satellites to many more dark
 matter models are required to ascertain the precise boost to the
 probability provided by lower concentration haloes. We also note that the stripping method has been calibrated to just one WDM model, the 3.3~keV thermal relic. This model is similar to our least extreme model, LA8,  and may not be appropriate for the other two models. We expect that these models will experience even more stripping than we predict here, pushing the preferred halo mass still higher.
 
\begin{figure}
     \includegraphics[scale=0.33]{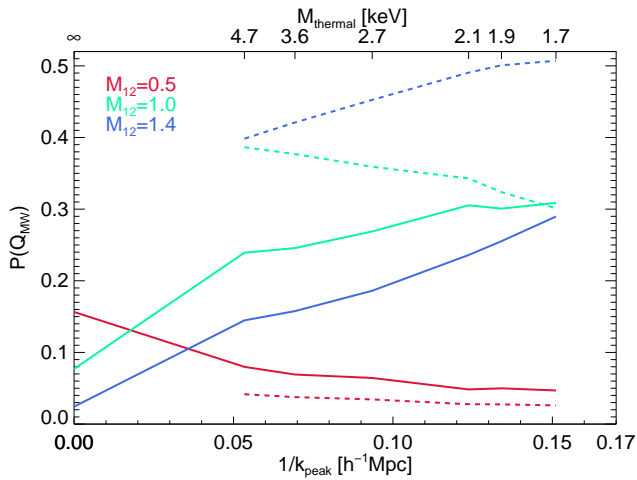}
 
     \caption{The probability that a Milky Way-like satellite \vmax
       distribution is drawn from the simulated distributions as a
       function of the inverse of the dimensionless matter power
       spectrum peak wave number, $k_\rmn{peak}^{-1}$. The \vmax
       function selection is made using the luminosity-ordered +
       baryonic effects correction scheme, and the galaxy formation
       model is LC16 with recalibration for the sterile neutrino
       models. The thermal relic masses corresponding to the value of
       $k_\rmn{peak}^{-1}$ for each of our models are displayed on the
       top axis. The values 1.7, 1.9, 2.1, 2.7, and 3.6~keV correspond
       to the 7~keV sterile neutrino with $L_6$=120, 18, 14, 10, and 8
       respectively; the model at $M_\rmn{thermal}=$4.7~keV is a
       10~keV sterile neutrino with $L_6=7$. We do not include
       $L_6=12$ (LA12) in this plot due to a lack of space. 
 The colours correspond to different host halo
       masses $0.5\times10^{12}\Msun$ (red), $1.0\times10^{12}\Msun$
       (green), and $1.4\times10^{12}\Msun$ (blue). Solid lines denote
       results calculated when the observed values of \vmax are
       derived from CDM simulations, and dashed where the
       \citet{WangMY15} factor is applied. }
     \label{PPWk}
 \end{figure}

 We end this subsection with a study of the probability of hosting a Milky Way-like \vmax function as a function of the dark matter power spectrum cutoff. We parameterize each of our models using the position of the peak of each matter power spectrum, which we denote $k_\rmn{peak}$. For CDM this value is formally infinite, therefore we consider the inverse of the peak,  $k_\rmn{peak}^{-1}$. We consider three halo masses ($0.5,1.0,1.4\times10^{12}\Msun$) and six sterile neutrino models 
 (7~keV, $L_6$=120, 18, 14, 10, and 8, plus 10~keV, $L_6=7$), and plot the results in Fig.~\ref{PPWk}. In order to make the connection to particle physics experiments and previous work on the subject, we also include equivalent thermal relic masses for our models on the top x-axis. These are the thermal relic masses that have the same value of $k_\rmn{peak}$ as our sterile neutrino models, with their matter power spectra calculated using the procedure of \citet{Viel05}.
 
 The value of the probability, $P(>Q_\rmn{MW})$, correlates with $k_\rmn{peak}^{-1}$ for all three halo masses. For the two more massive haloes, the trend is positive as a reflection of the suppression of the  \vmax function with $k_\rmn{peak}^{-1}$, for the lightest halo the trend is reversed due to over-suppression. The probability may increase by as much as a factor of three when the  \citet{WangMY15} factor is applied. However, we reiterate that this correction is based of just one WDM model and one satellite. A precise prediction will require a fit for every satellite with every model of interest, which we defer to later work.

 \section{Conclusions}
\label{con}	

The central densities of satellites have been the subject of much
recent study. Observations have been used to estimate the masses of
satellite galaxies and simulations have improved sufficiently to make
robust predictions for satellite density profiles. The observations
were found by \citet{BoylanKolchin11,BoylanKolchin12} to be discrepant
with N-body (dark matter only) simulations, which overpredict the
inner densities measured for the brightest Milky Way satellites. This
issue became known as the `too big to fail' problem.

Many solutions have been suggested, and in some cases they complement
one another. These include: assuming a relatively low mass for the
halo of the Milky Way \citep{Wang12,Cautun14b}; changing the
cosmological parameters \citep{Polisensky14}; the creation of a
central core by supernova feedback \citep{Brooks14}; a reduction in
the value of \vmax, reflecting lower halo growth induced by early mass
loss \citep{Sawala16a}; assuming that the dark matter is
self-interacting \citep{Zavala13}, that it couples to
radiation~\citep{Schewtschenko15}, or that it consists of sterile
neutrinos \citep{Lovell12}, in which cases satellite halos are less
dense that in the standard CDM model.

In this study we considered three of these possible solutions, namely
a low Milky Way halo mass, the suppression of \vmax by baryonic
effects, and sterile neutrino dark matter. In addition we considered
the impact of selecting satellites by stellar mass or luminosity
rather than by halo mass, as is done in an N-body simulation. Each
possibility was considered separately and in concert in order to
establish which combination of factors would provide the best match to
the measured Milky Way satellite \vmax function.

We computed Milky Way luminosity and \vmax functions for fourteen
Milky Way halo masses in the range $[0.5,1.8]\times10^{12}\Msun$ using
a modification of the \citet{Lacey16} version of the {\sc galform} 
semi-analytic galaxy formation model, described in Lacey et al. (2016), that was adapted to be run assuming an underlying Planck cosmology, PCH halo merger trees, and the
subhalo stripping algorithm introduced by \citet{JiangF15}. The dark
matter subhaloes were populated with galaxies by {\sc galform} and we
calculated the suppression of \vmax by baryonic effects using the
parametrisation introduced by \citet{Sawala16a}. We recalibrated the
semi-analytic stripping model against the CDM COCO simulation, and
recovered a good match between the PCH and N-body \vmax functions for
\vmax$\ge20$~\kms.

The sterile neutrino model was a 7~keV mass particle, chosen to be
consistent with the decay interpretation of the otherwise unexplained
3.55~keV line signal detected in clusters of galaxies and in M31
\citep{Bulbul14,Boyarsky14a}.The measured flux from these targets
implies a mixing angle for the sterile neutrino in the range
$\sin^2(2\theta)=[2,20]\times10^{-11}$. This corresponds to a
generation lepton asymmetry approximately in the range $L_6$=[8,12].
The value of the lepton asymmetry plays a role in setting the
free-streaming length; therefore we adopted three values of $L_6$: 8,
10, and 12. $L_6=8$ has the shortest free-streaming length and
$L_6=12$ the longest of the models we consider. For each combination
of these three sterile neutrino models, and for CDM, with the chosen
halo masses we generated $5000$ merger trees in order to take account
of the stochastic scatter introduced by different merger histories.

We showed that the models predict luminosity functions that tend to be 
steeper than, but still consistent with the data, even in the
luminosity range in which the satellite census is thought to be
complete (Figure.~\ref{LF}). Models that predict the correct number of
$M_V=-10$ galaxies produce LMC-like satellites in less than
10~per~cent of realisations. The suppression at low luminosities in
the sterile neutrino models leads to even better agreement with the observed
luminosity function.

A similar pattern was found in the \vmax functions, in that models
that host Magellanic Cloud analogues tend to overpredict the number of
less massive satellites unless they have a rather small total mass
(Figure.~\ref{VF}). As found by \cite{Sawala16b}, this tension is
eliminated when the suppression of \vmax by baryonic effects, which
decreases the median number of Milky Way satellites with
\vmax$>20$~\kms from 16 to 12, is taken into account. The agreement
with observations is better still for the sterile neutrino models,
especially since the lower concentrations of sterile neutrino haloes
translate into a lower host halo \vmax.

In order to determine how likely the Milky Way \vmax function is to
have been drawn from our PCH-generated \vmax distributions, we
characterise the variation between individual halo realisations using
the $Q$ statistic introduced by \citet{JiangF15}. The probability that
the Milky Way satellite \vmax function could be drawn from that
distribution is then $P(>Q_\rmn{MW})$, where $Q_\rmn{MW}$ is the value
of $Q$ for the Milky Way satellite \vmax function relative to the
simulated version. We find that, for halo masses
$\ge1\times10^{12}\Msun$, the selection of the brightest subhaloes
rather than all luminous subhaloes can increase $P(>Q_\rmn{MW})$ by a
factor of 10, and the correction of \vmax due to for baryonic effects
by up to a further factor of two (Fig.~\ref{PP}). Sterile neutrino
models have a higher likelihood than CDM models, and $P(>Q_\rmn{MW})$
is correlated with the free-streaming length. This trend is reversed
for smaller halo masses, due to the lack of massive satellites in the
sterile neutrino models.

We have thus shown that satellite \vmax functions like that of the
Milky Way are generated in the CDM cosmology. They are more common in
sterile neutrino cosmologies, which allow for satellites to reside in
more massive haloes, of which there are fewer still than in CDM. This
model also has the attraction that it matches the data for a set of
sterile neutrino parameters that account for the 3.5~keV line feature.

There are many uncertainties that remain in the sterile neutrino
calcuation presented in this paper. The first is that the exact degree
of halo tidal stripping is unknown; in principle this needs to be
assesed using simulations for each sterile neutrino case
\citep{Bozek16}. It also remains to be seen whether these sterile
neutrino models generate enough faint ($M_V>-8$) satellites
\citep{Lovell16,Schneider16} or match the Lyman-$\alpha$ forest flux measurements (\citealp{Viel13,Schneider14}; but see also \citealp{Garzilli15}). There is currently enough uncertainty in
both the galaxy formation model and the observational constraints that
these models cannot be ruled out; however, tighter constraints from
future observational surveys may help establish if these models are
viable.

With respect to the CDM cosmology, we find that Milky Way-like systems are rare but by no means impossible. This represents a refinement on other studies, such as \citet{Sawala16a} and \citet{Fattahi16}, that similarly find the Milky Way systems in high resolution simulations typically underpredict the satellite \vmax functions but not to a severe degree. More stringent tests of the model will be realised as both observations and simulations attain the ability to measure the velocity dispersions of faint galaxies: if the census of the twenty brightest satellites is already complete then the disagreement with observations is very strong \citep{JiangF15}. Hints in this direction have been discovered in a new set of fairly bright but extended satellites of the Milky Way \citep{Caldwell16} and M31 \citep{Collins13} in which the measured velocity dispersions may be lower than CDM allows for.

 \section*{Acknowledgements}

 We would like to thank Mikko Laine for supplying the code that
 calculates the sterile neutrino distribution functions,
 Wojtek Hellwing for use of the COCO simulation data, and Carlton Baugh for the use of the {\em Planck}-tuned {\sc galform} models. We would also like to thank John Helly for providing the subhalo accretion times in {\sc galform}. CSF acknowledges
 ERC Advanced Grant 267291 'COSMIWAY.' This work used
 the DiRAC Data Centric system at Durham University, operated by the
 Institute for Computational Cosmology on behalf of the STFC DiRAC HPC
 Facility (www.dirac.ac.uk). This equipment was funded by BIS National
 E-infrastructure capital grant ST/K00042X/1, STFC capital grant
 ST/H008519/1, and STFC DiRAC Operations grant ST/K003267/1 and Durham
 University. DiRAC is part of the National E-Infrastructure. This work
 is part of the D-ITP consortium, a programme of the Netherlands
 Organization for Scientific Research (NWO) that is funded by the
 Dutch Ministry of Education, Culture and Science (OCW). This work was
 supported in part by an STFC rolling grant to the ICC . This project has received funding from the European Research Council (ERC) under the European Union's Horizon 2020 research and innovation programme (GA N$^{\circ}$ 694896).

  \bibliographystyle{mnras}
 % \bibliography{bibtex}%, astro,combined_numsm}

\bsp
  \label{lastpage}
  
  \appendix
  \section{Change in galaxy formation model}
  \label{app}
  
\begin{figure}
     \includegraphics[scale=0.5]{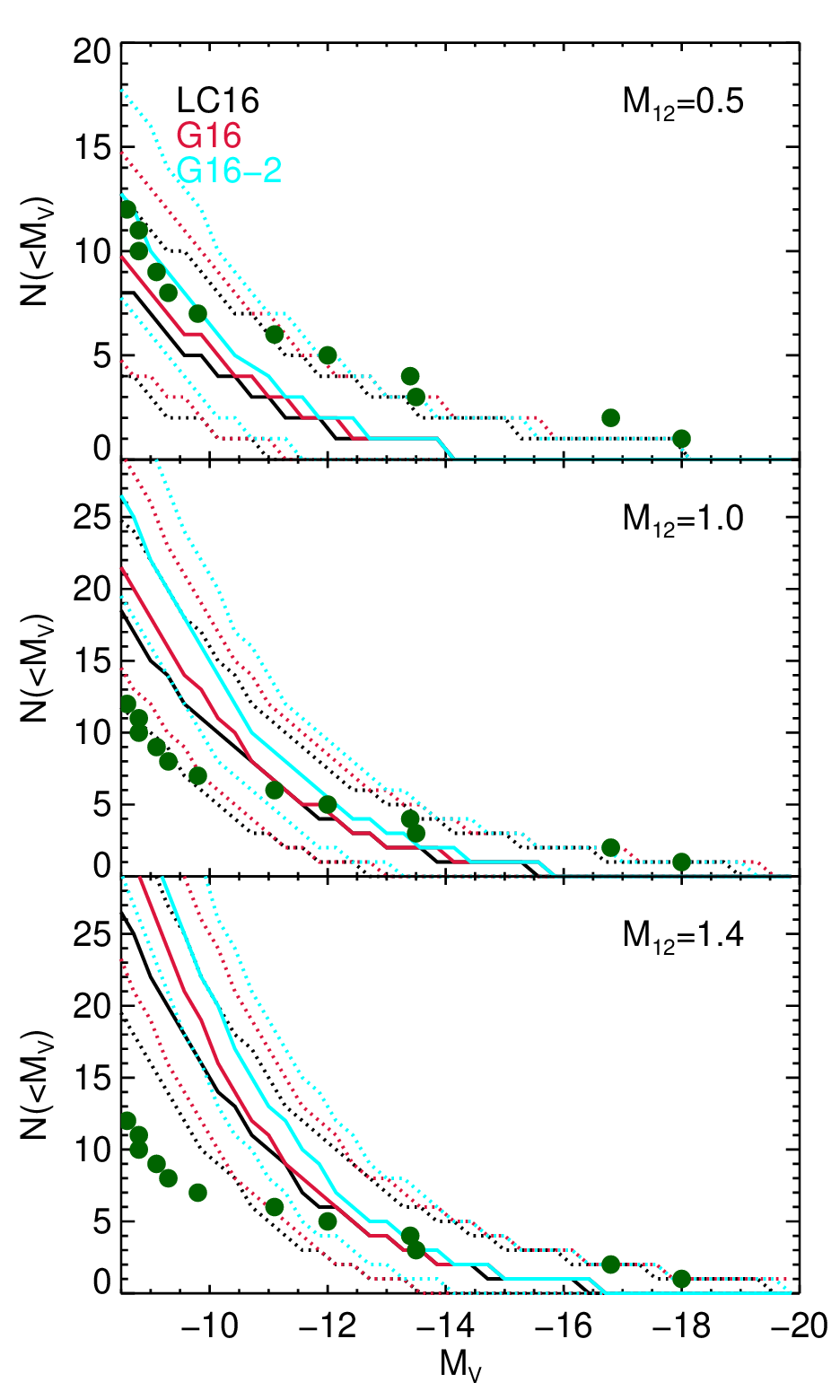}
     \caption{Cumulative satellite luminosity function for three
       $\Lambda$CDM 
       galaxy formation models and three halo masses. The solid lines
       denote the median number of satellites
       brighter than $M_{V}$ across all of our realisations, and the
       dotted curves mark the 9 and 95 percentiles. Each model is
       denoted by a different colour: LC16 (black), G16 (red) and
       G16-2 (cyan) as indicated in the legend. The top, middle, and
       bottom panels show the mass functions for central haloes of
       mass 0.5, 1.0, and $1.4\times10^{12}\Msun$ respectively. The
       circles mark the observed Milky Way satellite luminosity
       function.}
     \label{LF3}
 \end{figure}

  The galaxy formation model can influence the satellite luminosity
  function in two ways: regulating the luminosity of the galaxy that
  can be formed in a halo with a given formation history, and
  destroying satellites through mergers with the host galaxy. In this
  Appendix we discuss these effects by analysing 
  three versions of the {\sc galform} model: LC16, G16, and G16-2. We start
  with luminosity functions for our  three assumed 
  halo masses ($0.5,1.0,1.4\times10^{12}\Msun$). As seen in
  Fig.~\ref{LF3}, the main difference
 amongst these models is that G16 produces more faint galaxies than
 LC16 due to
 its weaker feedback and G16-2 even more due to the lower merger rates
 of satellites onto the main galaxy (see below). As a result, the
 satellite luminosity functions are steeper than in LC16.

\begin{figure}
     \includegraphics[scale=0.5]{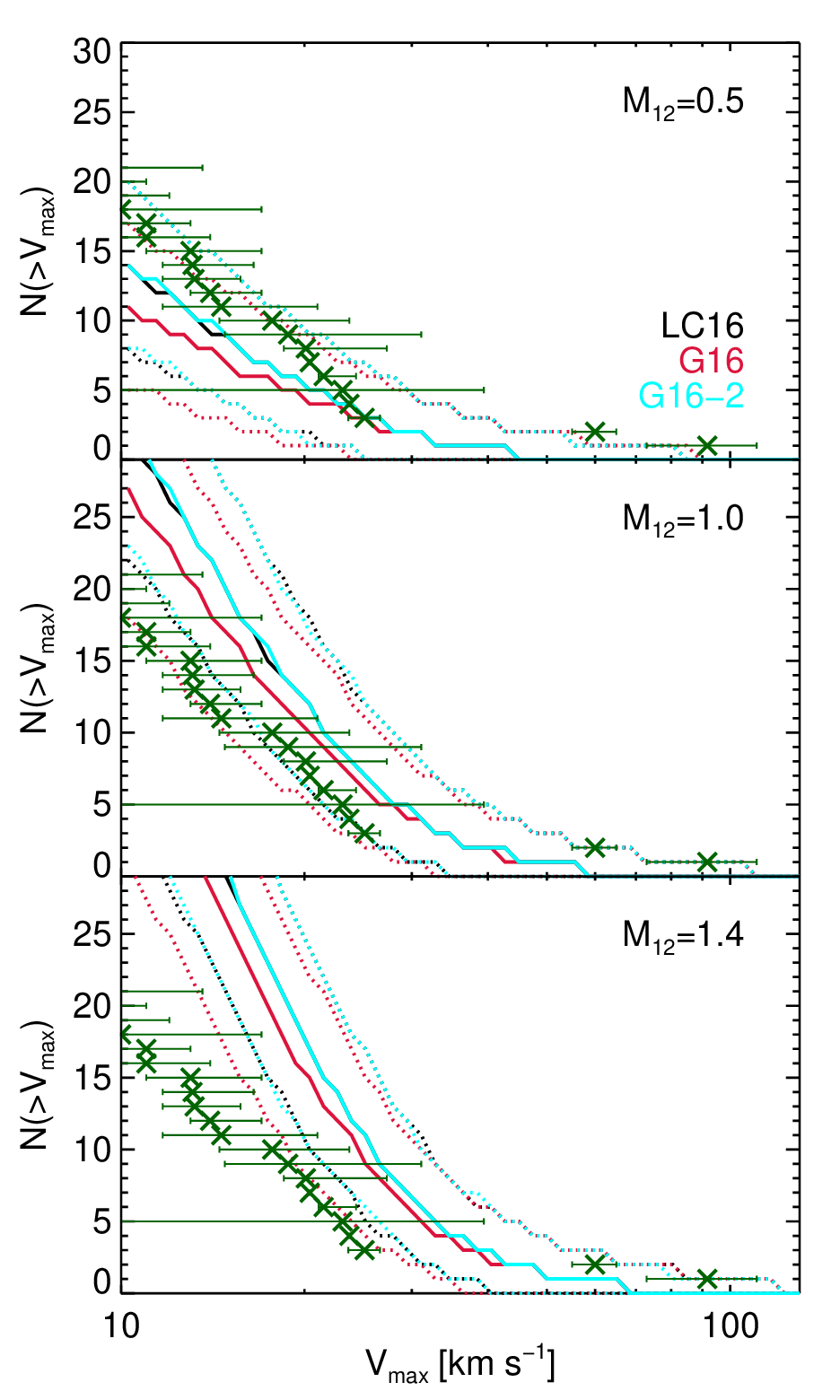}
     \caption{Cumulative satellite \vmax function for the CDM versions
       of LC16 (black), G16 (red), and G16-2 (cyan). We include all
       luminous satellites, and have applied the baryonic physics
       feedback correction from equation~\ref{eqn:torus}. The solid
       lines denote the median \vmax across all of our realisations,
       and the dotted curves again mark the 5 and 90 percentiles. The
       top, middle, and bottom panels again show the mass functions
       for central haloes of mass 0.5, 1.0, and
       $1.4\times10^{12}\Msun$ respectively. The dark green crosses
       mark the inferred Milky Way satellite \vmax function assuming
       CDM.}
     \label{VFA}
 \end{figure}

 The \vmax functions are plotted in Fig.~\ref{VFA}. LC16 and G16-2
 give nearly identical results, but for G16 the \vmax function is
 slightly suppressed suggesting that haloes are more readily destroyed in this model. This
 explains the relative amplitudes of the luminosity functions in
 Fig.~\ref{LF3}: LC16 suppresses the luminosity function through
 stronger feedback, and G16 through higher merger rates.
  
\begin{figure}
     \includegraphics[scale=0.33]{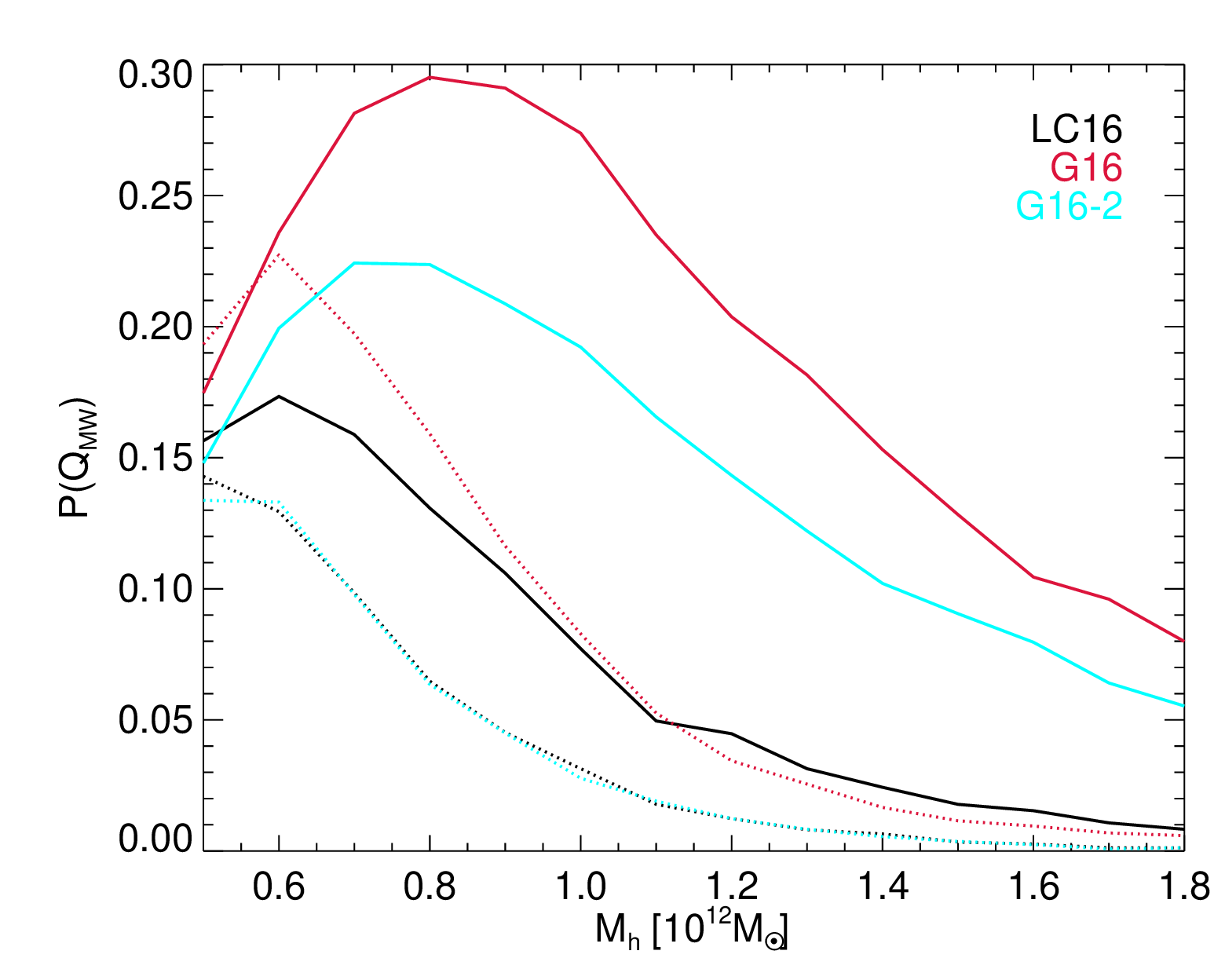}
 
     \caption{The probability that a Milky Way-like satellite \vmax
       distribution is drawn from the simulated distributions as a
       function of halo mass. The models used are LC16 (black), G16
       (red) and G16-2 (cyan). Dotted lines denote results calculated
       from the base model (no luminosity-ordering or feedback
       suppression), and solid where luminosity-ordering and feedback
       suppression are included. }
     \label{PPA}
 \end{figure}

 We now consider the implications for the probability of retrieving
 the Milky Way satellite luminosity function from the three model
 \vmax distributions. We calculate $P(>Q_\rmn{MW})$ as a function of
 halo mass and plot the results in Fig.~\ref{PPA}. The base models (no
 luminosity-ordering or feedback suppression) of LC16 and G16-2 are
 almost identical, which results directly from their near-identical
 \vmax functions. The elimination of some large satellites increases
 the amplitude of the G16 probability curve by up to a factor of two
 relative to G16-2. The boost to the likelihood introduced by
 including luminosity-ordering information is stronger for the G16 and
 G16-2 models, since their relatively low mass satellites can form
 galaxies potentially as bright as their more massive counterparts.
 For G16
 this model can generate Milky Way-like satellite \vmax functions in
 up to 30~per~cent of realizations for ow halo masses. We
 therefore conclude that greater scatter in the halo mass-luminosity
 relation may also play a role in removing the too big to fail
 problem. However, the models must be able to increase the scatter
 whilst
 simultaneously making all satellites fainter in order to fit the
 Milky Way satellite luminosity function.
  
\end{document}